# Impact of Deployment on Energy Efficiency of Sub-THz Transmission

Hardy Halbauer, Le Hang Nguyen, Thorsten Wild
Nokia Bell Labs
Stuttgart, Germany
{hardy.halbauer, le_hang.nguyen, thorsten.wild}@nokia-bell-labs.com

*Abstract*— Sub-THz bands are promising high bandwidth and data rates, and in the recent years the device technologies made large progress and provided a multitude of transceiver, power amplifier (PA) and phased array devices supporting the frequency bands above 100 GHz. The more painful aspect of sub-THz transmission is the increased power consumption, caused by the large data rates and the related data conversion and processing effort, and on the analog side the low achievable PA efficiency and the reduced achievable output power. When planning a deployment of sub-THz communication systems, the target coverage and throughput can be achieved with a variety of scenarios, which will be different with respect to locations and number of base stations and system architectures. Although leading to similar performance, they will differ significantly in the overall power consumption. With an accurate power consumption model, including also baseband (BB) processing functionality, and system level simulations for different hybrid beamforming and MIMO schemes the related variations in power consumption in relation to a given performance are evaluated. This paper shows the critical design aspects for energy efficient sub-THz deployments by highlighting the sub-THz specific trade-offs between different number of BS with different transmit powers but also changing number of BB units and RF chains.

I. INTRODUCTION

Potential use cases discussed for sub-THz applications comprise large indoor scenarios like shopping malls, airport halls, office areas or conference rooms [1]. To provide sufficient coverage and throughput for a highly varying number of users in the targeted coverage area, multiple access points or base stations are needed. This leads to a trade-off between performance and overall power consumption. Transmission in sub-THz bands with high bandwidth is facing specific challenges. The increased path loss and the technological limitations of reduced output power of PAs can be overcome e.g. with high gain beam forming. Other aspects are the non-linearity leading to high output power backoff, and the low PA efficiency of the PAs operating at frequencies above 100 GHz, causing high power consumption. Further, due to the high carrier frequency the phase noise of oscillators is increased and impacts the overall performance. Those aspects can be addressed with optimized waveforms [2], [5]. AI-based methods allow to "learn" waveforms and constellation shapes with reduced PAPR to achieve higher average output power and having increased robustness against phase noise [3], [6]. Including neural network (NN) based processing in the receiver further improves performance, but adds additional complexity and increases power consumption due to the high processing rate. This leads to the question how to best provide coverage and good performance with lowest power consumption, i.e. how to achieve best energy efficiency (EE) for a specific target deployment and target performance, in terms of e.g: achieving a specific minimum throughput within a given deployment area. From [4] it is obvious that high gain requirements ask for a high number of antenna elements. But since [2] reveals that the expected number of reasonably feasible beams and MIMO streams would be only in the range of 1 up to 4, a number of RF chains, much lower than the number of antenna elements, would be sufficient. This enables hybrid beamforming with reduced number of RF chains and MIMO streams, thus reducing power consumption in BB processing and RF chains. Our contribution in this paper therefore focuses on low complexity hybrid beamforming approaches with up to four MIMO streams and up to 32 user equipments (UEs) within the considered exemplary deployment areas. The power consumption for achieving a specific area throughput using different number of BSs and hybrid beamforming architectures is analyzed for different numbers of simultaneously served UEs. For this a detailed power consumption model has been derived, including also baseband (BB) processing parts. This allows for assessing the sub-THz specific energy efficiency trade-offs between a smaller number of BS with higher transmit power compared to more BS with lower transmit power, but overall higher number of BB units and RF chains, serving the same deployment area.

The paper is structured as follows: First, the considered deployment scenarios along with the transmitter and receiver architectures are introduced, followed by performance analysis results of different hybrid architectures.. Then a power consumption model is derived, comprising all relevant building blocks from BB processing, data conversion, analog up/down conversion and RF blocks for PA, LNA and beamforming. To use practical data of different technologies and designs the models for the analog components and data converters are derived from recently published device data and adapted or extrapolated to different output powers and bandwidths. The methodology for deriving the model for BB processing is taken from literature [15][16][17][18] and extensively modified and adapted to the current system design and related functional blocks of processing. It includes all digital transmitter (Tx) and receiver (Rx) functions, also reflecting the impact of receiver variants for AI optimized waveforms. Finally, the performance versus related power consumption is analyzed and discussed.

II. DEPLOYMENT SCENARIO

*A. System simulation outline*

The following analysis assumes an indoor scenario with an extension of 30x60 m, which exemplarily can represent e.g. a shopping mall or large office room. Two cases with 4 and 8 deployed base stations (BS) are considered in the analysis (Fig. 1). The BS height is 4 m above ground, and the BSs are pointing towards the inner part of the area with a fixed mechanical downtilt of 21 degrees. The BSs are equipped with a uniform planar antenna array of 8 x 8 half-lambda spaced elements, which can be used either as one large array, or partitioned into 2 subpanels of size 4 x 8 elements or 4 subpanels of size 4 x 4 elements (Fig. 2). Each subpanel is driven by a separate RF chain which transmits one individual MIMO stream, thus being able to serve 1, 2 or 4 users simultaneously, respectively.. But since the total transmit power per BS is kept constant, the transmit power per subpanel is 3 dB and 6 dB less for the 2- and 4-subpanel cases, respectively. In contrast to the usually applied system simulation evaluation our analysis follows a different approach. It evaluates the total throughput within the deployed area versus the different BS deployments and transmit power settings. This allows to provide a direct comparison of totally consumed power for providing similar performance within the considered area.

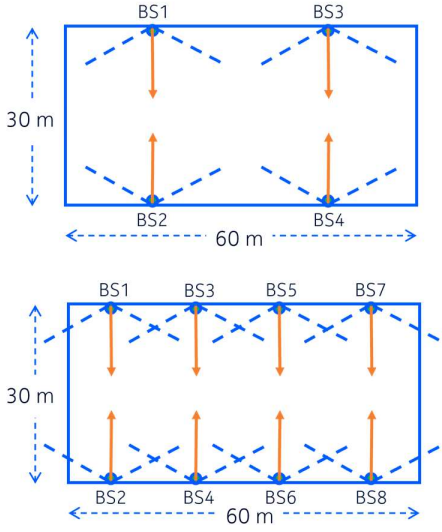

Fig. 1: Functional blocks of RF chain with phased array (Tx)

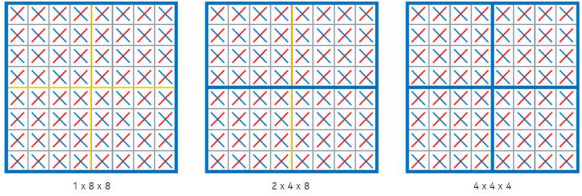

Fig. 2: Array configurations: 1 x 8 x 8, 2 x 4 x 8 and 4 x 4 x 4

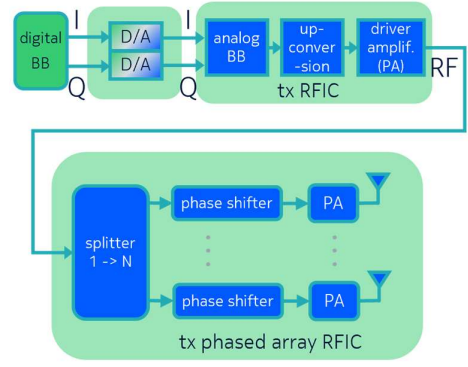

Fig. 3 Functional blocks of RF chain with phased array (Tx)

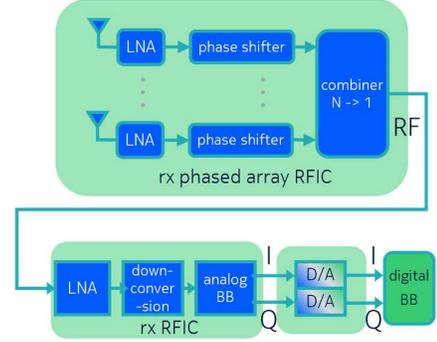

Fig. 4: Functional blocks of RF chain with phased array (Rx)

UEs are assumed to be dropped every second randomly within the deployment area and assigned to the BS achieving best throughput. If more than one UE per RF chain is dropped, multiple UEs share one BS and are scheduled on this RF chain within this one second period according to a proportional fair algorithm. Based on the achievable throughput and the time share of the scheduling the throughput at the BS is calculated. If less UEs than available RF chains are dropped, some of these RF chains remain unused and do not contribute to the overall power consumption within the deployment area. For the present analysis 1000 of such drops are taken into account.

Initial investigations with 2 subpanels with 4 x 8 elements showed slightly better performance in this specific scenario than the "tall" 8 x 4 subpanels, probably because it allows smaller beams and therefore better interference suppression in the horizontal direction. Therefore the analysis only considers the 4 x 8 configuration. Hybrid beamforming is realized with a grid-of-beam (GoB) approach, for each subpanel a beam out of a GoB of 8 x 8 beams can be selected. To reflect the limited elevation angles of the scenario the beams in vertical direction cover an angle range of +/- 15° relative to the mechanical downtilt; in horizontal direction an angle range of +/- 60° relative to boresight is covered.

### B. System architectures

A common way of implementing Tx and Rx devices for sub-THz bands is the realization as highly integrated RFICs, comprising all functional blocks from analog BB processing, up- and downconversion, PA and LNA. Sub-THz phased arrays are often implemented as a combination of splitter, analog phase shifters or vector modulators, followed by a PA per antenna element on Tx side, or one LNA per antenna element, phase shifter or vector modulator and combiner on Rx side. A waveguide connects phased array and Tx or Rx RFIC. In recent publications, devices with similar architectures but different transmit output power levels have been reported, including DC power consumption values, e.g. [8], [10], [11], [12]. For BB processing the number of functional blocks scales with the number of MIMO streams. The number of DACs and ADCs is two per RF chain, if we assume quadrature I and Q BB signals connected to the BB interface of the RFICs.

Fig. 3 and Fig. 4 show the investigated architectures on Tx and Rx side, for one RF chain, highlighting the groups of functional blocks reflecting the power consumption model structure used in the following. At lower frequencies on Tx side often one high power PA per RF chain is driving splitter, analog phase shifters and multiple antenna elements and needs to overcome the related losses [14]. In sub-THz bands, due to the physical limitations of the hardware, one PA per antenna element at lower power is more efficient, or enables only the realization. On Rx side a similar structure, here with highly integrated LNAs and phase shifters per antenna element, is considered for the analysis.

For multiple RF chains multiple of these architectures will be applied. If more than one MIMO stream should be applied, the precoding will become part of the digital processing, feeding then multiple RF chains. However, due to complexity reasons in this paper we consider only straightforward variants with one RF chain and one MIMO stream per subpanel. Extension towards more complex structures is straightforward if the additional BB processing blocks and the power consumption models are adapted accordingly.

*C. Performance analysis*

With system simulations the average throughput within the deployment area in downlink has been evaluated for the two exemplary deployment scenarios of Fig. 1 and the array configurations of Fig. 2. Parameter were the total Tx power per BS, varying from 11 dBm up to 35 dBm in steps of 6 dB, and the number of simultaneously active UEs within the area. As channel model an adapted version of the 3GPP channel model for indoor hotspots, originally valid up to 100 GHz [13], has been applied. Measurements at 140 GHz [2] revealed that the propagation properties are quite close to that model, if the number of reflectors and number of reflected paths are adapted according to the measurement results. Some further simulation parameters are given in TABLE I.

TABLE I. SYSTEM SIMULATION PARAMETERS

| Parameter | Value |
|---|---|
| Carrier frequency | 140 GHz |
| Signal bandwidth | 5 GHz |
| Channel model | 3GPP TR 38.901 InH, adapted to 140 GHz |
| No. of BS | 4 and 8 |
| Antenna array | Uniform planar array, 8 x 8 elements, vertically polarized, λ/2 spaced, element HPBW = 65° |
| Antenna element | HPBW = 65° in horizontal and vertical direction, max. gain = 8 dBi |
| Subpanels | 1 x 8x8 / 2 x 4x8 / 4 x 4x4 |
| Beamforming | 8 x 8 grid-of-beams |
| Angular range of grid-of-beams | Horizontally: +/- 60° rel. to boresight Vertically: +/- 15° rel. to mech. downtilt |
| Scheduler | Proportional fair, one beam per MIMO layer |
| No. of UEs | 4, 8, 16, 32 / 2 x-polarized antennas |
| UE antenna | 2 omnidirectional elements, x-polarized, max. gain = 0 dBi |
| UE rx noise figure | 9 dB |

Area throughput results for the 30m x 60m deployment area are shown in Fig. 5 - Fig. 7 for different subpanel configurations, 4 … 32 simultaneously active UEs and 4 or 8 BS deployment. For each total Tx power level per BS value the left 4 bars show the area throughput for 4 BS, the right 4 bars represent the area throughput of the 8 BS scenario. Throughput is calculated as average throughput over all drops of UEs within the simulation duration, as explained in detail in section II.A. A general outcome is that for all subpanel configurations the 4-BS scenarios have lower throughput than the 8-BS scenarios, because the distances to the BS is for a larger number of locations higher than in the 8-BS case and therefore more UEs are dropped in more noise-limited parts of the area, even if the Tx power of the BS is higher.

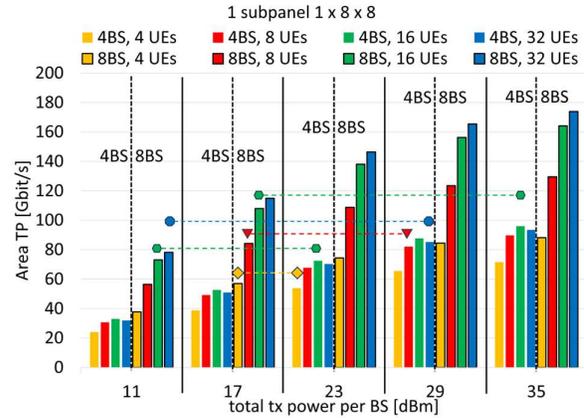

Fig. 5. Total throughput in deployment area for 1 subpanel 1x8x8

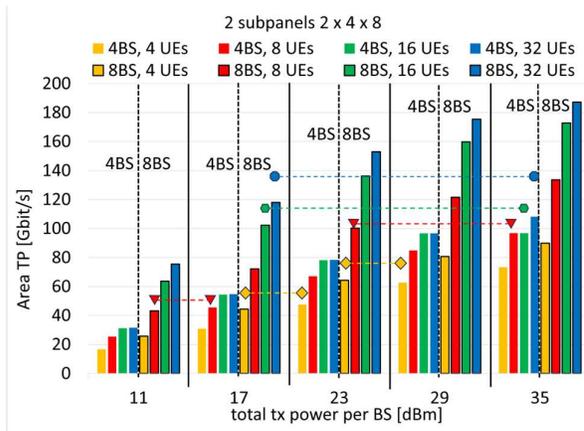

Fig. 6. Total throughput in deployment area for 2 subpanels 2x4x8

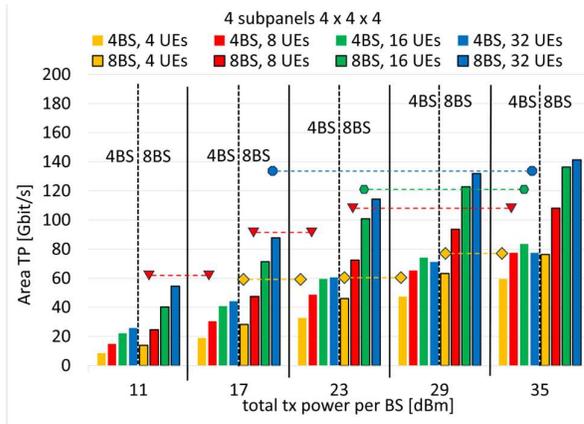

Fig. 7. Total throughput in deployment area for 4 subpanels 4x4x4

The results lead further to the observation that, if 8 BS are deployed, a similar throughput as with 4 BS can be achieved at a much lower Tx power level per BS. In Fig. 5 for example 16 UEs served with 4 BS at 23 dBm total Tx power show a similar throughput as 16 UEs served by 8 BS at only 11 dBm (left green dashed line). Also for 8 UEs 4 BS with 29 dBm Tx power per BS and 8 BS with 17 dBm the same throughput results (red dashed line). Further examples in Fig. 5 (shown by yellow, blue and a further green dashed line) are 4 UEs / 4 BS@ 23 dBm vs. 8 BS@17 dBm, 32 UEs / 4 BS@29 dBm vs. 8 BS@11 dBm, or 16 UEs / 4 BS@35 dBm vs. 8 BS@17 dBm. Similar observations hold for the other subpanel configurations in Fig. 6 and Fig. 7. So, for each number of simultaneously served UEs, a similar throughput as with 4 BSs can be achieved with 8 BSs, but at high Tx output power savings per BS in the range of 6 – 12 dB, in some cases even up to 18 dB. In linear scale this is factor 4 – 16 or even up to 64. The cell edge performance, not shown here, behaves similar to the total throughput when applying the same Tx output power reduction.

If we assume a PA efficiency of 10% the DC power consumption of the PA is factor of 10 higher, so that the DC power saving is significant, even if the number of BSs is increased by a factor of 2. So the overall power saving is higher than the increase of the number of BS. It is clear that, if a specific throughput is targeted, the way of deployment has significant impact on the energy efficiency of the whole scenario. But it is not only the higher number of PAs to be considered, because the base stations contain also other functional blocks which will contribute to power consumption. This makes it worth to analyze the potential power savings in more detail. In fact, it is required to explore a trade-off between less overall transmit power per BS and the increasing number of RF chains with BB processing, data converters and analog circuits which comes with increased no. of BS, different hybrid architectures and related RF chains. Therefore, a detailed power consumption model has been derived and will be introduced in Section III. It can be easily scaled to the different scenarios and no. of RF chains involved.

## III. POWER CONSUMPTION MODEL

Communication system functions for frequency bands in the sub-THz range are often realized as fully integrated RFICs. This is due to the fact that the PA output power is limited, and interfaces between the different building blocks will cause additional losses and further reduce overall efficiency. Therefore, the power consumption model used for the following analysis is based on the architectures shown in Fig. 3 and Fig. 4. On Tx side the essential blocks are the BB processing, digital-to-analog conversion, Tx RFIC comprising the analog BB block, a direct upconverter and a driver PA, which can drive either a phased array unit or directly a horn antenna. Finally, a Tx phased array comprising a splitter, controllable phase shifter or vector modulator and a PA per antenna element. On Rx side the corresponding functional blocks are a phased array with LNA and a controllable phase shifter or vector modulator per antenna element, followed by a combiner, an Rx RFIC including LNA, downconverter and analog BB processing, digital-to-analog converters and the BB processing block.

In the following, for each analog functional block the derivation of power consumption is explained, and the parameters according to which the power consumption is scaled. For this we rely on recently published data of existing sub-THz and D-band devices and adapt towards the parameters of the considered system. In this simplified model we assume one MIMO stream and one RF chain per subpanel. Calculation of more complex MIMO structures is straightforward by scaling the number of blocks and power consumption values of these blocks accordingly.

The digital BB processing blocks shown in Fig. 8 are modelled following the method of [17], [18] which has been adapted to higher frequency bands and technology progress as proposed in [15], [16]. The computational complexity, i.e. the number of complex operations is measured as giga-floating-point operations per second (GFLOPS). With the intrinsic efficiency $E_{intr}$ given in GFLOPS/W and adapted to technology progress according to [16] the power consumption is calculated.

### A. Digital BB processing

In this paper, the complexity of LDPC encoder and decoder, constellation mapping and network and control processing are taken from [17] and [18], all other values are counted according to the detailed functionality as implemented in our specific system and described in detail in [7]. In the following the used complexity formulas are given in detail. The complexity values of each functionality can vary differently with one or more of the involved design parameters. Major parameters are spectral efficiency of modulation scheme $s$ (for QPSK, 16QAM, 64QAM and 256QAM $s$ = 2, 4, 6 and 8, respectively), number of spatial streams $K$, RF chains $M$, antenna elements per subpanel $N_M$, no. of subpanels per BS $N_{SP}$, constellation sampling rate $fc$, oversampling rate $fs$ (both given in Hz), FFT size $n_{FFT}$, filter length $n_{taps}$. The principle is to calculate the number of complex operations per sample or symbol (depending on functional block) and multiply with the number of samples $fc$ or symbols $fs$ per second and with the number of streams $K$ or RF chains $M$. Dividing this value by the intrinsic efficiency yields the power consumption of this functional block. The typical values derived below for each block are for a signal bandwidth of 5 GHz, with the roll-off of the single carrier signal of 0.3 this leads to a symbol frequency $fc$ of 3.93216 GHz. The used parameters and derived complexities in the following, reflect typical values of the system assumed in the analysis ($K$ = 1 assumed) .

#### 1) LDPC encoding and decoding complexity

$$C_{LDPC\_enc} = \frac{dc}{dsc \cdot n_{bits\_enc}} \cdot s \cdot K \cdot fc \quad (1)$$

$$C_{LDPC\_dec} = \frac{n_{it} \cdot n_{op\_real}}{sc_{compl} \cdot n_{bits\_dec}} \cdot s \cdot K \cdot fc \quad (2)$$

The formulas were taken from [17], with inclusion of some parameters instead of using numbers directly to better adapt to the configured LDPC codes. With the additional parameters check node degree of LDPC code $dc$=14, implementation specific scaling factor assuming binary operations $dsc$=8, no. of simultaneously encoded bits $n_{bits\_enc}$=3, no. of decoding iterations $n_{it}$=50, no. of real operations per bit $n_{op\_real}$=35, scaling to complex operation $sc_{compl}$=2, no. of simultaneously decoded bits $n_{bits\_dec}$=3, the encoding complexity $C_{LDPC\_enc}$ is 4.6 up to 18.4 GFLOPS and the decoding complexity $C_{LDPC\_dec}$ is 2293 up to 9175 GFLOPS for $s$ = 2 up to 8, respectively.

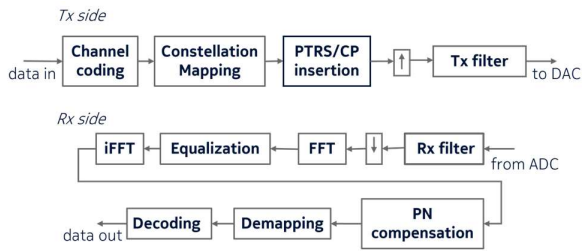

Fig. 8. Functional blocks of BB processing

*2) Constellation symbol mapping, Cyclic Prefix (CP) and Phase Tracking Reference Signal (PTRS) insertion*

$$C_{mapping} = s^{1.5} \cdot K \cdot fc \quad (3)$$

$$C_{PTRS\_CP\_ins} = \frac{n_{CP} + n_{PTRS}}{n_{block}} \cdot n_{op\_ins} \cdot K \cdot fc \quad (4)$$

The formula for $C_{mapping}$ is taken from [17], for $C_{PTRS\_CP\_ins}$ it is derived from our implementation in [7]. With cyclic prefix length $n_{CP}$=288 symbols, no. of PTRS symbols $n_{PTRS}$=128, no. of operations per symbol insertion $n_{op\_ins}$=1, FFT block length $n_{block}$=4384 for $C_{Mapping}$ results a complexity of 11 up to 89 GFLOPS (for $s$ = 2 up to 8) and for $C_{PTRS\_CP\_ins}$ of 0.4 GFLOPS.

*3) Tx and Rx filter*

$$C_{filter} = n_{op\_filt} \cdot n_{taps} \cdot M \cdot fs \quad (5)$$

For Tx and Rx a similar FIR filter can be assumed, but the values of no. of operations per filter tap $n_{op\_filt}$ and no. of filter taps $n_{taps}$ may also be different. Typical values are $n_{op\_filt}$=2 and $n_{taps}$=40, resulting in $C_{filter}$ = 1258 GFLOPS.

*4) FFT and IFFT*

FFT and IFFT have similar complexity. The formulas of [17] are applied, but keeping the parameters no. of stages $n_{stages}$ (typ. = 13 for 212-FFT), no. of operations per butterfly structure $n_{op\_bfl}$ (typ. =3) and no. of symbols calculated per butterfly $n_{sym\_bfl}$ (typ. =2) generic. With the typical values this results in $C_{FFT}$ = 76.7.

$$C_{FFT} = \frac{n_{stages} \cdot n_{op\_bfl}}{n_{sym\_bfl}} \cdot K \cdot fc \quad (6)$$

*5) Channel estimation and equalization*

Channel estimation complexity depends on the ratio of DMRS symbols per equalization block $r_{DMRS}$. The ratio of data blocks per equalization block is $r_{data}$. For the increased complexity of complex division, a scaling factor $n_{op\_factor}$ has been included. Equalization complexity strongly depends on the selected equalization algorithm. Two options are considered. For zero-forcing (ZF) $n_{op\_factor}=n_{op\_ZF}$ = is 1, for the minimum mean square error (MMSE) algorithm, due to required matrix operations $n_{op\_factor}=n_{op\_MMSE}$ is in the order of 2 $n_{FFT}$.

$$C_{est} = r_{DMRS} \cdot n_{op\_factor} \cdot K \cdot fc \quad (7)$$

$$C_{Eq\_ZF} = r_{data} \cdot n_{op\_ZF} \cdot K \cdot fc \quad (8)$$

$$C_{Eq\_MMSE} = r_{data} \cdot n_{op\_MMSE} \cdot K \cdot fc \quad (9)$$

Typical values for the described scenario are $C_{Est}$=0.4, $C_{Eq\_ZF}$ = 3.7 and $C_{Eq\_MMSE}$ = 29908 GFLOPS.

*6) Phase noise compensation*

Phase noise compensation in time domain uses $n_{op\_interp}$ interpolations between the $n_{PTRS}$ PTRS symbols included in the data block of $n_{block}$ symbols. Complexity is related to the no. of PTRS symbols and the no. of interpolated correction values between the PTRS locations. For the implemented version in [7] we get

$$C_{PN} = \left((n_{PTRS} - 1) \cdot n_{op\_interp} + (n_{block} - n_{PTRS} - n_{CP}) \cdot n_{comp\_sample}\right) \cdot \frac{n_{dbps}}{n_{bps} \cdot n_{block}} \cdot K \cdot fc \quad (10)$$

$n_{comp\_sample}$ denotes the number of operations per sample for PN compensation, $n_{dbps}$ and $n_{bps}$ denote no. of data blocks and total no. of blocks within an FFT block, respectively. With $n_{op\_interp}$=4, $n_{comp\_sample}$=1, $n_{dbps}$=13, $n_{bps}$=14 a value of $C_{PN}$ = 4 results.

*7) Demapping*

Two different demapping variants have been analyzed [7]. The MaxLogMap demapper (Demap_mlm) shows increasing complexity with constellation size. The neural network based demapper (Demap_NN) has higher complexity already for QPSK, but lower increase of complexity with increasing constellation size. The related complexity for the MaxLogMap demapper is

$$C_{Demap\_mlm} = (4 \cdot 2^s - 2) \cdot s \cdot K \cdot fc \quad (11)$$

With the typical values introduced above this yields the complexities $C_{Demap\_mlm}$=110 GFLOPS for $s$=2 (QPSK) and 5992 GFLOPS for $s$ = 6 (64QAM).

The complexity for the NN demapper can be calculated as

$$C_{Demap\_NN} = \left((2 \cdot n_{L1} + 1) \cdot n_{L2} + (2 \cdot n_{L2} + 2) \cdot n_{L3} + (2 \cdot n_{L3} + 1) \cdot n_{Lout}\right) \cdot K \cdot fc \quad (12)$$

with $n_{L1}$, $n_{L2}$, $n_{L3}$ and $n_{Lout}$ denoting the number of neurons in layer 1, 2, 3 and output layer, respectively. $n_{Lout}$ is identical to $s$ since it represents the no. of bits per symbol. With assumed values of 2, 64, 64 and $s$ for these parameters the complexity for QPSK is 8898 GFLOPS, for 64QAM it is 9414 GFLOPS.

*8) Network and control processing*

Complexity values for these blocks representing platform control and network processing for downlink and uplink parts, $C_{Net\_\&\_CTRL\_DL}$, $C_{Net\_\&\_CTRL\_UL}$. The scaling coefficients and exponents for bandwidth, spectral efficiency $s$ and number of streams $K$ have been taken from [18] and scaled to the parameter values of the considered system. Most important, the resulting scaling factor of 250 is due to the higher bandwidth of the present system. For completeness the complexity values used in this analysis for $K$=1 are given in TABLE II.

TABLE II. COMPLEXITY OF NETWORK AND CONTROL PROCESSING

| $s$ | 2 | 4 | 6 | 8 |
|---|---|---|---|---|
| $C_{Net\_\&\_CTRL\_DL}$ [GFLOPS] | 669 | 1336 | 2002 | 2669 |

| $C_{Net\_\&\_CTRL\_UL}$ [GFLOPS] | 443 | 884 | 1326 | 1768 |

Finally, for data transfer activities (memory and register access) an overhead factor $OV$ of 1.5 on the total BB processing complexity values is considered. With this, the total BB power consumption is calculated from the formulas above and the intrinsic efficiency $E_{intr}$, already taking into account $K$ RF chains, $M$ subpanels and modulation scheme $s$:

$$P_{BB\_Tx} = (C_{LDPC\_enc} + C_{mapping} + C_{PTRS\_CP\_ins} + C_{filter} + C_{Net\_\&\_CTRL\_DL}) \cdot OV \cdot E_{intr} \quad (13)$$

$$P_{BB\_Rx} = (C_{filter} + 2 \cdot C_{FFT} + C_{Eq} + C_{PN} + C_{Demap} + C_{LDPC\_dec} + C_{Net\_\&\_CTRL\_UL}) \cdot OV \cdot E_{intr} \quad (14)$$

with $C_{Eq} = C_{EqZF}$ or $C_{Eq\_MMSE}$, $C_{Demap} = C_{Demap\_mlm}$ or $C_{Demap\_NN}$.

### B. DA- and AD-converter (DAC / ADC)

For the power consumption $P_{DAC}$ of DACs there is no generically applicable derivation depending on sampling rate and resolution. Scanning literature for devices providing sampling rates above 10 Gsps and resolution of 16 bits show that the power consumption of most devices is in the range of 800 mW up to 2.4 W, but also lower and higher values for specific design targets or technologies are reported. For the power consumption model therefore an average value of 1.25 W per DAC has been assumed. This value can be adapted if specific devices, designs or technologies are envisaged. Examples for the power values can be found e.g. in Table I of [19], or in data sheets of suppliers, e.g. [20].

For ADCs the similar method as in [4], now based on the latest update of ADC data in [21], is applied. The ADC power consumption $P_{ADC}$ is modelled with the envelope Walden Figure of Merit (FOM), which gives a sampling frequency and technology progress dependent normalized minimum energy in femto-Joule per conversion step. For 10 Gsps the current value is 6.2 fJ/conv.-step. The power consumption for a given sampling frequency and resolution is calculated according to

$$P_{ADC}(fs) = FOM(fs) \cdot fs \cdot 2^{ENOB} \quad (15)$$

With $fs$ = sampling frequency and $ENOB$ = effective number of bits. With a sampling frequency of 10 Gsps and a resolution of 12 bits this leads to 180 mW per ADC, for 14 bits it consumes already 720 mW

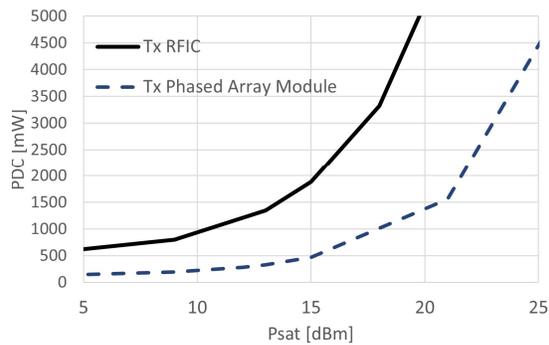

Fig. 9. Power consumption vs saturated output power of Tx RFIC and Tx phased array module

### C. Tx and Rx RFIC

This building block contains the RF chain from the BB to the RF signal, including analog BB processing, up- or downconversion and a driver amplifier on Tx and an LNA on Rx side. Critical point on Tx side is a sufficiently high output power, so that the signal can be split up for the phase shifter chains of the phased array. Various real devices with different Tx power capabilities have been published [8], [12], [22], so that a power consumption model adaptable for different Tx output powers $P_{RFIC\_Tx}$ can be derived which considers almost fixed contribution of BB, LO and mixer part, but output power dependent contribution to DC power consumption. On Rx side there is no significant dependency on system parameters, if we consider mainly D-band and high bandwidth systems. Related references providing power consumption data show low variation if the whole Rx RFIC is considered. The references [1], [8], [12], [23] suggest power consumption values in similar ranges for this block. With these data the typical power consumption $P_{RFIC\_Tx}$ derived for Tx side is shown on Fig. 9.

For the Rx side based on the published data a typical fixed value of $P_{RFIC\_Rx}$ = 960 mW is assumed.

### D. Phased array

The Tx and Rx phased array according to Fig. 4 operates on RF only, on Tx side with a splitter and per antenna element a controllable vector modulator, and a PA (Tx). On Rx side the corresponding blocks are an LNA and a controllable vector modulator per antenna element, and a combiner. Only the PA power consumption is depending on the Tx output power, the other blocks have fixed power consumption, but their number scales with the number of antenna elements. Model values have been derived from published data e.g. in [8], [10], [11], [12]. The resulting power consumption $P_{Ph\_Tx}$ versus saturated output power of the Tx part of the module is also shown in Fig. 9. For the Rx part a $P_{Ph\_Rx}$ of 240 mW is assumed.

### E. Waveform and BB processing impact, power supply

Besides scaling according to the different system parameters also the impact of the modulation on Tx and Rx processing must be considered. Further, the power supply efficiency $E_{PS}$ of 92%, resulting in an overall increase of power consumption by 8% [18] and the duty cycle ($DC_{DL}$ = 75% downlink and $DC_{UL}$ = 25% uplink) are included.

The digital BB processing scales with the no. of data streams $K$. In our consideration this means with the product of the no. of subpanels $M$ per BS and the total no. of BS $N_{BS}$ within the deployment area. The same for the DACs and ADCs, increased by a factor of 2 for I and Q channel BB processing. Tx and Rx RFICs also scale with $M$. For each phased array, the no. of modules (vector modulator and PA) scales with the number of antenna elements per phased array $N_M$, the no. of phased arrays scales with $M$. In addition to these considerations, for each system variant the power consumption of the Tx part varies according to Tx output power, the power consumption of the BB processing varies with the waveform, the modulation scheme and the Rx processing (either standard or NN demapper, equalizer algorithm). With this we have the power consumption model complete:

$$P_{Tx} = \left[\left[(P_{BB\_Tx} + (2 \cdot P_{DAC} + P_{RFIC\_Tx} + N_M \cdot P_{Ph\_Tx}) \cdot M\right) \cdot DC_{Tx}\right] \cdot \left(1 + (1 - E_{PS})\right)\right] \cdot N_{BS} \quad (16)$$

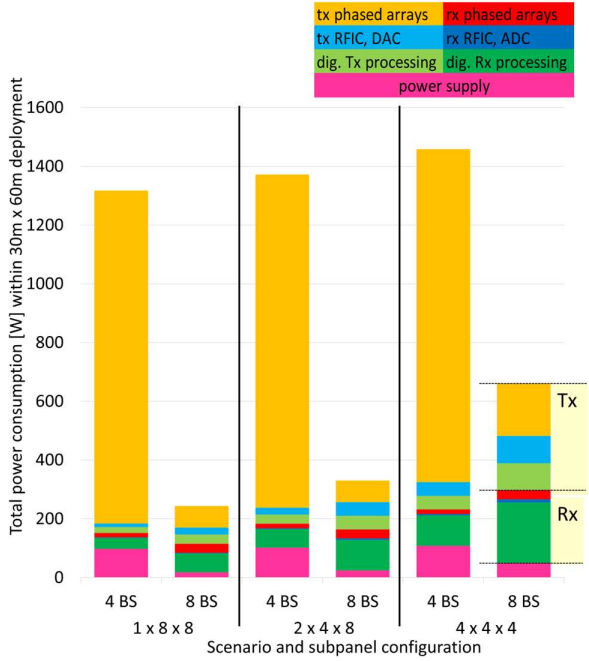

Fig. 10. Total power consumption within deployment area

$$P_{Rx} = \left[\left[\left(P_{BB\_Rx} + \left(2 \cdot P_{ADC} + P_{RFIC\_Rx} + N_M \cdot P_{Ph\_Rx}\right) \cdot M\right) \cdot DC_{Rx}\right] \cdot \left(1 + (1 - E_{PS})\right)\right] \cdot N_{BS} \quad (17)$$

where the dependency on selected algorithms, channel bandwidth, sampling rates and modulation schemes of the BB processing blocks, and their scaling according to the no. of MIMO streams is already included in the related formulas (1) … (14).

IV. POWER CONSUMPTION OF DIFFERENT SCENARIOS

In this section the power consumption of some exemplary cases described in section II.C is compared. Out of the huge amount of system simulation data for each subpanel configuration and for 16 simultaneously served UEs the cases are compared, where 4 BS at maximum Tx power of 35 dBm per BS reach less or similar throughput in the deployment area than 8 BS with less Tx power per BS. These are the cases shown in TABLE III.

TABLE III. 4BS AND 8BS CASES FOR COMPARISON

| Subpanel | Tx power per BS [dBm] | |
|---|---|---|
| | 4 BS | 8 BS |
| 1 x 8 x 8 | 35 | 17 |
| 2 x 4 x 8 | 35 | 17 |
| 4 x 4 x 4 | 35 | 23 |

Fig. 10 shows the total power consumption within the deployment area, either served with 4 or 8 BS. It also shows the contribution of the different functional blocks according to Fig. 3 and Fig. 4. At the bottom of the columns the power supply is modelled as a percentage of the total power consumption. Throughput in the deployment area is comparable within same subpanel configuration. On receive side a ZF channel estimator and a NN demapper as described in [7] is applied. Thus, regarding total power consumption per deployment area using 8 BS instead of 4 BS, with 1/2/4 subpanels we observe roughly a factor 6 / 5 / 2 reduction, which is significant.

V. DISCUSSION OF RESULTS, CONCLUSION AND OUTLOOK

This exemplary analysis shows that for sub-THz deployments the transmitter part can become the dominating contribution to the power consumption of the system, especially if high Tx output power is required. With increased number of BS the required Tx output power per BS can be significantly reduced while maintaining the area coverage and throughput on the same level. However, there is a trade-off coming from the increased number of BB processing blocks due to the higher number of RF chains. For the 8 BS case of 2x2 subpanels the digital processing part already consumes more power than the Tx phased array part. Nevertheless, overall power consumption is still less than for the 4-BS case. It is clear that the described scenarios are only exemplary, and additional trade-offs not yet explored are possible, e.g. using different antenna pattern or MIMO arrangements, or reduced resolution of ADCs compensated with increasing number of antenna elements or subpanels. Also the type of deployment and traffic load may be different or dynamically changing. The main conclusion of this work is that with the specific properties of sub-THz devices and the high signal bandwidth and related data rates the way of deployment matters.

Ongoing and future research work has to focus on a further joint detailed analysis of power consumption, energy efficiency and target system level performance for different deployment scenarios, model parameter variations and trade-offs mentioned above. For this a flexible and detailed power consumption model is now available, which covers a wide range of different functional blocks, especially also the digital signal processing, and can be easily parametrized to reflect different scenarios, types of devices and design variants. A regular revisit of the technological progress of sub-THz RFIC hardware and system design as well as signal processing capabilities, and adaptation of the essential model parameters is mandatory. In combination with related system level simulations this will ensure accurate assessments, also for future upcoming systems and use cases.

VI. ACKNOWLEDGEMENT

This work is partially funded by the Hexa-X-II project, receiving funding from the Smart Networks and Services Joint Undertaking (SNS JU) under the European Union's Horizon Europe research and innovation programme under Grant Agreement No 101095759.

It is further partially funded by the Federal Ministry of Research, Technology and Space (BMFTR), Germany under the project "ESSENCE-6GM" (grant number: 16KISK162K).

VII. REFERENCES

[1] Hexa-X, "Deliverable D2.3: Radio models and enabling techniques towards ultra-high data rate links and capacity in 6G," 2023. [Online]. Available: https://hexa-x.eu/wp-content/uploads/2023/04/Hexa-X-D2_3_v1.0.pdf (use cases)
[2] Hexa-X, "Deliverable D2.2: Initial radio models and analysis towards ultra-high data rate links in 6G," 2021. [Online]. Available: https://hexa-x.eu/wp-content/uploads/2022/01/Hexa-X-D2_2.pdf (channel model)

[3] Hexa-X-II, "Deliverable D4.5: Final results of 6G radio key enablers," 2025. [Online]. Available: https://hexa-x-ii.eu/wp-content/uploads/2025/03/Hexa-X-II_D4_5_v1_edit.pdf

[4] H. Halbauer and T. Wild, "Towards power efficient 6G sub-THz transmission," in Proc. European Conference on Networks and Communications (EuCNC) & 6G Summit, 2021.

[5] D. Marasinghe, L. H. Nguyen, J. Mohammadi, Y. Chen, T. Wild, and N. Rajatheva, "Constellation shaping under phase noise impairment for sub-THz communications," in Proc. IEEE International Conference on Communications (ICC), 2024.

[6] D. Marasinghe, L.H. Nguyen, J. Mohammadi, Y. Chen, T. Wild and N. Rajatheva, "Waveform learning under phase noise impairment for sub-THz communications," IEEE Transactions on Communications, 2024

[7] L. H. Nguyen, H. Heimpel, D. Marasinghe, H. Halbauer, and T. Wild, Sub-THz waveform evaluation in the D-band: A proof of concept study," in Proc. European Conference on Networks and Communications (EuCNC) & 6G Summit, 2024.

[8] A. Singh, M. Sayginer, M. J. Holyoak, J. Weiner, J. Kimionis, M. Elkhouly, Y. Baeyens, and S. Shahramian, "A D-Band Radio-on-Glass Module for Spectrally-Efficient and Low-Cost Wireless Backhaul," in 2020 IEEE Radio Frequency Integrated Circuits Symposium (RFIC)

[9] M. Elkhouly, M. J. Holyoak, D. Hendry, M. Zierdt, A. Singh, M. Sayginer, S. Shahramian, and Y. Baeyens, "D-band Phased-Array TX and RX Front Ends Utilizing Radio-on-Glass Technology," in 2020 IEEE Radio Frequency Integrated Circuits Symposium (RFIC), 2020

[10] M. Elkhouly et al., "Fully Integrated 2D Scalable TX/RX Chipset for D-Band Phased-Array-on-Glass Modules," 2022 IEEE International Solid-State Circuits Conference (ISSCC), San Francisco, CA, USA

[11] A. Ahmed, L. Li, M. Jung, S. Li, D. Baltimas and G. M. Rebeiz, "140-GHz 2-D Scalable On-Grid 8× 8-Element Transmit–Receive Phased Arrays With Up/Down Converters Demonstrating a 5.2-m Link at 16 Gbps," in IEEE Transactions on Microwave Theory and Techniques, vol. 72, no. 5, May 2024

[12] A. Karakuzulu, W. A. Ahmad, D. Kissinger and A. Malignaggi, "A Four-Channel Bidirectional D-Band Phased-Array Transceiver for 200 Gb/s 6G Wireless Communications in a 130-nm BiCMOS Technology," in IEEE Journal of Solid-State Circuits, vol. 58, no. 5, pp. 1310-1322, May 2023

[13] 3GPP TR 38.901, "Study on channel model for frequencies from 0.5 to 100 GHz," v18.0.0, Technical Report, Mar. 2024.

[14] S. Wesemann, J. Du, and H. Viswanathan, "Energy efficient extreme MIMO: Design goals and directions," IEEE Communications Magazine, Oct. 2023.

[15] C. Desset, N. Collaert, S. Sinha and G. Gramegna, "InP / CMOS co-integration for energy efficient sub-THz communication systems," 2021 IEEE Globecom Workshops (GC Wkshps), Madrid, Spain, 2021

[16] C. Desset, P. Wambacq, Y. Zhang, M. Ingels, and A. Bourdoux, "A flexible power model for mm-wave and THz high-throughput communication systems," In PIMRC Workshop on Enabling Technologies for Terahertz Communications (ETTCOM), London, UK, August 2020.

[17] C. Desset and B. Debaillie, "Massive MIMO for energy-efficient communications," in EuMC, London, UK, Oct. 2016.

[18] C. Desset, B. Debaillie and F. Louagie, "Modeling the hardware power consumption of large scale antenna systems," 2014 IEEE Online Conference on Green Communications (OnlineGreenComm), (Online Only) AZ, USA, 2014

[19] H. Huang, "Ultra-high-speed digital-to analog converter for optical communications", Dissertation, Universität Stuttgart, [Online]. Available: https://elib.uni-stuttgart.de/items/096217dc-b576-478b-b7d0-3d0b749d8686

[20] Analog Devices, data sheet AD9176, [Online]. Available: AD9176 (Rev. B)

[21] B. Murmann, "ADC Performance Survey 1997-2024," [Online]. Available: https://github.com/bmurmann/ADC-survey.

[22] M. Sayginer et al., "A 110-170 GHz Phase-Invariant Variable-Gain Power Amplifier Module with 20-22 dBm Psat and 30 dBm OIP3 Utilizing SiGe HBT RFICs," 2023 IEEE Radio Frequency Integrated Circuits Symposium (RFIC), San Diego, CA, USA, 2023

[23] T. Maiwald et al., "A Full D-Band Low Noise Amplifier in 130 nm SiGe BiCMOS using Zero-Ohm Transmission Lines," 2020 15th European Microwave Integrated Circuits Conference (EuMIC), Utrecht, Netherlands, 2021

[24] A. Karakuzulu, M. H. Eissa, D. Kissinger and A. Malignaggi, "Full D-Band Transmit–Receive Module for Phased Array Systems in 130-nm SiGe BiCMOS," in IEEE Solid-State Circuits Letters, vol. 4